\documentclass[aps,preprint,showpacs]{revtex4}
\usepackage{amsmath}
\usepackage{graphicx}

\begin{document}

\title{Small damping approach in Fermi-liquid theory}
\author{V.M.~Kolomietz$^{1,2}$, S.V.~Lukyanov$^{1}$ and S.~Shlomo$^2$}
\affiliation{$^{1}$Institute for Nuclear Research, Prosp. Nauky 47, 03680 Kyiv, Ukraine\\
$^{2}$Cyclotron Institute, Texas A\&M University, College Station, Texas
77643, USA}

\begin{abstract}
The validity of small damping approximation (SDA) for the quasi-classical
description of the averaged properties of nuclei at high temperatures is
studied within the framework of collisional kinetic theory. The isoscalar
collective quadrupole vibrations in hot nuclei are considered. We show
that the extension of the SDA, by accounting for the damping of the
distribution function $\delta f$ in the collision integral reduces the
rate of variation with temperature of the Fermi surface distortion effects.
The damping of the $\delta f$ in the collision integral increases significantly
the collisional width of the giant quadrupole resonance (GQR) for small
enough values of the relaxation time. The temperature dependence of the
eigenenergy of the GQR becomes much more weaker than in the corresponding
SDA case.
\end{abstract}

\maketitle

\section{Introduction}

In this work, we study the damping of nuclear multipole vibrations within
the semiclassical kinetic theory. Semiclassical approaches are quite
instructive for the description of averaged properties and macroscopic
collective motion in multiparticle systems, see Ref. \cite{kosh04}\ and
references therein. In many cases, they allow us to obtain analytical
results and represent them in a transparent way. However a significant
simplification is achieved in these approaches due to the so-called small
damping approximation (SDA), which assumes that the relaxation rate is small
enough compared to the energy of the collective excitation. This
approximation is reasonable in a cold Fermi system where the excitation
energy is high enough (due to the distortion of the Fermi surface) and the
relaxation rate is quite small (zero sound regime) \cite%
{lipi93,bape91,abkh59}. With the heating of the Fermi system the energy of
the collective vibrations decreases and the two-body relaxation rate
increases simultaneously (zero- to first- sound transition) \cite{lipi93}.
Thus with increasing temperature the ratio of the relaxation rate to the
excitation energy increases rapidly and the use of the SDA can be invalid.
In this case the analytical results for eigenenergy, width,
strength function, etc., obtained with the SDA, have to be verified
independently.

In this paper we study the validity of the SDA within the Fermi liquid
theory considering the excitation of the giant quadrupole resonance (GQR) in
a hot nucleus. In Sect. II we derive the equations of motion for the
collective quadrupole vibrations in a finite heated Fermi liquid drop for
both the SDA and the general case. In Sect. III we provide a numerical study
of the temperature dependence of the excitation energy and the relaxation
rate for the above mentioned cases. Our conclusions are given in Sect.~IV.

\section{Equations of motion}

To derive the equations of motion for the nuclear shape variables, we start
from the following kinetic equation for the small deviation $\delta f$ of
the distribution function $f\equiv f(\mathbf{r},\mathbf{p};t)$ from the one
in equilibrium, $f_{\mathrm{eq}}$, taking into account the collision
integral $\delta \mathrm{St}$ linearized with respect to equilibrium \cite%
{abkh59,kota81},
\begin{equation}
{\frac{\partial }{\partial t}}\delta f+\hat{L}\delta f=\delta \mathrm{St}%
[\delta f].  \label{2.1}
\end{equation}%
On the left hand side of Eq. (\ref{2.1}), the operator $\hat{L}$ represents
the drift terms including the selfconsistent mean field $V$,
\begin{equation*}
\hat{L}\delta f={\frac{\mathbf{p}}{m}}\cdot \mathbf{\nabla }_{\mathbf{r}%
}\delta f-\mathbf{\nabla }_{\mathbf{r}}V_{\mathrm{eq}}\cdot \mathbf{\nabla }%
_{\mathbf{p}}\delta f-\mathbf{\nabla }_{\mathbf{r}}\delta V\cdot \mathbf{%
\nabla }_{\mathbf{p}}f_{\mathrm{eq}}\,.
\end{equation*}%
We will follow the nuclear fluid dynamic approach, see Ref. \cite{kosh04},
and take into account the dynamic Fermi surface distortion up to
multipolarity $l=2$,
\begin{equation}
\delta f=-\left( {{\frac{\partial f}{{\partial \epsilon }}}}\right) _{%
\mathrm{eq}}\ \sum_{l,m_{l}}^{l=2}\delta f_{lm_{l}}(\mathbf{r},t)Y_{lm_{l}}(%
\hat{p}).  \label{2.2}
\end{equation}%
Here $\epsilon $ is the quasiparticle energy \cite{abkh59}. A generalization
of this approach to the case of an arbitrary multipolarity $l$ of the Fermi
surface distortion can be done in a straightforward way, see Ref. \cite%
{koma92}. Using Eqs. (\ref{2.1}) and (\ref{2.2}) we will derive a closed set
of equations for the following $\mathbf{p}$-moments of the distribution
function, namely, local particle density $\rho $, velocity field $u_{\nu }$
and pressure tensor $P_{\nu \mu }$, in the form (for details, see Refs. \cite%
{kota81,kopl95,kiko96}) of

\begin{equation}
{\frac{\partial \delta \rho }{\partial t}}+{\frac{\partial }{\partial r_{\nu
}}}(\rho _{\mathrm{eq}}u_{\nu })=0,  \label{sys1}
\end{equation}%
\begin{equation}
m\rho _{\mathrm{eq}}\frac{\partial u_{\alpha }}{\partial t}+\rho _{\mathrm{eq%
}}\frac{\partial }{\partial r_{\alpha }}\left( \frac{\delta ^{2}\mathcal{E}}{%
\delta \rho ^{2}}\right) _{\mathrm{eq}}\delta \rho +{\frac{\partial
P_{\alpha \nu }^{\prime }}{\partial r_{\nu }}}=0,  \label{sys2}
\end{equation}%
\begin{equation}
{\frac{\partial P_{\alpha \beta }^{\prime }}{\partial t}}+P_{\mathrm{eq}%
}\left( \frac{\partial u_{\alpha }}{\partial r_{\beta }}+\frac{\partial
u_{\beta }}{\partial r_{\alpha }}-\frac{2}{3}\delta _{\alpha \beta }\frac{%
\partial u_{\nu }}{\partial r_{\nu }}\right) =\delta \mathrm{St}_{\alpha
\beta }.  \label{sys3}
\end{equation}%
Here $\mathcal{E}$ is the internal energy density, which is the sum of the
kinetic energy density of the Fermi motion and the potential energy density
associated with the nucleon-nucleon interaction. The equilibrium pressure of
a Fermi gas, $P_{\mathrm{eq}}$, is given by
\begin{equation}
P_{\mathrm{eq}}={\frac{1}{3m}}\int {\frac{gd\mathbf{p}}{(2\pi \hbar )^{3}}\ }%
p^{2}f_{\mathrm{eq}},  \label{2.6}
\end{equation}%
$P_{\alpha \beta }^{\prime }$ is the deviation of the pressure tensor from
its isotropic part due to the Fermi surface distortion
\begin{equation}
P_{\alpha \beta }^{\prime }=-{\frac{1}{m}}\int {\frac{gd\mathbf{p}}{(2\pi
\hbar )^{3}}}(p_{\alpha }-mu_{\alpha })(p_{\beta }-mu_{\beta })\left( {\frac{%
\delta f}{\delta \epsilon }}\right) _{\mathrm{eq}}\sum_{m}\delta f_{2m}(%
\mathbf{r},t)Y_{2m}(\hat{p}),  \label{2.7}
\end{equation}%
$\delta \mathrm{St}_{\alpha \beta }$ is the second moment of the collision
integral
\begin{equation}
\delta \mathrm{St}_{\alpha \beta }={\frac{1}{m}}\int {\frac{gd\mathbf{p}}{%
(2\pi \hbar )^{3}}\ }p_{\alpha }p_{\beta }\delta \mathrm{St}[\delta f]
\label{2.8}
\end{equation}%
and $g=4$ is the spin-isospin degeneracy factor. In Eqs. (\ref{sys1})-(\ref%
{sys3}) and in the following expressions repeated Greek indices are to be
understood as summed over.

Using the Fourier transformation for the pressure
\begin{equation}
P_{\alpha \beta }^{\prime }(t)=\int {\frac{d\omega }{2\pi }}e^{-i\omega
t}P_{\alpha \beta ,\omega }^{\prime },  \label{2.10}
\end{equation}%
and similarly for the other time dependent variables, we find the solution
to Eq. (\ref{sys3}) as
\begin{equation}
P_{\alpha \beta ,\omega }^{\prime }={\frac{{i\omega }\tau }{{1-i\omega \tau }%
}}P_{\mathrm{eq}}\Lambda _{\alpha \beta ,\omega },  \label{2.11}
\end{equation}%
where we used the symbol
\begin{equation}
\Lambda _{\alpha \beta ,\omega }=\nabla _{\alpha }\chi _{\beta ,\omega
}+\nabla _{\beta }\chi _{\alpha ,\omega }-{\frac{2}{3}}\delta _{\alpha \beta
}\nabla _{\lambda }\chi _{\lambda ,\omega }  \label{2.12}
\end{equation}%
for this combination of gradients of the Fourier transform $\chi _{\alpha
,\omega }$ of the displacement field. The time derivative of the
displacement field $\mathbf{\chi }(\mathbf{r},t)$ is defined as the velocity
field, hence
\begin{equation}
u_{\alpha ,\omega }=-i\omega \chi _{\alpha ,\omega }.  \label{2.13}
\end{equation}%
To obtain Eq. (\ref{2.11}) we have also used the fact that the tensor $%
\delta \mathrm{St}_{\alpha \beta }$, Eq. (\ref{2.8}), can be reduced to
\begin{equation}
\delta \mathrm{St}_{\alpha \beta ,\omega }=-{\frac{1}{\tau }}P_{\alpha \beta
,\omega }^{\prime },  \label{2.14}
\end{equation}%
due to our restriction to quadrupole deformation of the Fermi surface. This
is because the $l=0$ and $1$ components of the expansion (\ref{2.2}) do not
contribute to the collision integral, reflecting the conservation of
particle number and momentum in a collision, respectively. The relaxation
time $\tau $ in Eq. (\ref{2.14}) is derived by \cite{bert78}
\begin{equation}
\frac{1}{\tau }=-\frac{\int d\mathbf{p}\ p^{2}Y_{20}(\hat{p})\delta \mathrm{%
St}[\delta f]}{\int d\mathbf{p}\ p^{2}Y_{20}(\hat{p})\delta f}.  \label{t1}
\end{equation}

The collision integral $\delta \mathrm{St}[\delta f]$ is related to real
transitions in the interparticle collisions and the expression for this
collision integral should contain $\delta $-function associated with the
conservation of energy.\ Note that the form of Eq. (\ref{2.14}) is also
correct for a non-Markovian collision term, with the collision time $\tau $
being dependent on the frequency $\omega $, see Ref. \cite{kiko96}. This $%
\omega $-dependence is associated with the memory effects in the
interparticle collisions and the collision integral takes then the following
form \cite{koma92}
\begin{equation}
\delta \mathrm{St}[\delta f]=\int \frac{gd\mathbf{p}_{2}d\mathbf{p}_{3}d%
\mathbf{p}_{4}}{2(2\pi \hbar )^{6}}\ W(\{\mathbf{p}_{j}\})\ \delta (\Delta
\mathbf{p})\sum_{j=1}^{4}\left. \frac{\delta Q}{\delta f_{j}}\right\vert _{%
\mathrm{eq}}\delta f_{j}\ \left[ \delta (\Delta \epsilon +\hbar \omega
)+\delta (\Delta \epsilon -\hbar \omega )\right] .  \label{dst}
\end{equation}%
Here, $f_{i}\equiv f(\mathbf{r},\mathbf{p}_{i};t),$
\begin{equation*}
\Delta \epsilon =\epsilon _{1}+\epsilon _{2}-\epsilon _{3}-\epsilon
_{4},\quad \Delta \mathbf{p}=\mathbf{p}_{1}+\mathbf{p}_{2}-\mathbf{p}_{3}-%
\mathbf{p}_{4},
\end{equation*}%
where $\epsilon _{j}=p_{j}^{2}/2m+V_{\mathrm{eq}}(\mathbf{r})$ is the
classical single-particle energy, $Q$\ is the Pauli blocking factor
\begin{equation*}
Q=(1-f_{1})(1-f_{2})f_{3}f_{4}-f_{1}f_{2}(1-f_{3})(1-f_{4}),
\end{equation*}%
$W(\{\mathbf{p}_{j}\})=(d\sigma _{\mathrm{in}}/d\Omega )4g(2\pi \hbar
)^{3}/m^{2}$ is the probability of the scattering of nucleons near the Fermi
surface and $d\sigma _{\mathrm{in}}/d\Omega $ is the corresponding in-medium
differential cross section.

The collision integral of Eq. (\ref{dst}) provides the relaxation time $\tau
_{\mathrm{SDA}}$ of small damping approximation. It is given by \cite%
{koma92,ayik92}%
\begin{equation}
\tau _{\mathrm{SDA}}=\frac{\alpha \hbar }{T^{2}[1+\zeta (\hbar \omega /2\pi
T)^{2}]},  \label{t2}
\end{equation}%
where $T$ is the temperature and $\alpha $\ and $\zeta $\ are parameters.
The phenomenological parameter $\alpha $ is determined by the in-medium
spin-isospin averaged nucleon-nucleon cross section $\sigma _{\mathrm{in}}$%
.\ The magnitude of parameter $\zeta =1$ was earlier obtained by Landau \cite%
{land57} in the calculation of the absorption coefficient in a Fermi liquid.
The parameter $\zeta $ can be also evaluated in a direct way within the
framework of a kinetic theory if the exact expression for the collision
integral is known. In a general case, the $\omega $-dependence (memory
effect) in $\delta \mathrm{St}[\delta f]$ is caused by three sources \cite%
{kopl95}: (\textit{i}) variation of the distribution function $\delta f$, (%
\textit{ii}) variation of the mean field $\delta V$, and (\textit{iii})
screening effect for the two-body scattering in hot Fermi system due to high
frequency collective vibrations. For the sake of simplicity we will only
consider the contribution to $\delta \mathrm{St}[\delta f]$ from the
variation of the distribution function $\delta f$. The constant $\zeta $ is
then given by $\zeta =3$ \cite{koma92}.

The frequency $\omega $\ in Eq. (\ref{t2}) is assumed to be real. This
reflects the fact that we neglect the damping of the distribution function $%
\delta f$ in Eq. (\ref{dst}) due to the small damping approximation. In the
case of the complex eigenfrequency $\omega =\omega _{R}-i\omega _{I}$ with
non-vanishing imaginary part $\omega _{I}>0$ the delta function in Eq. (\ref%
{dst})\ should be spread. Thus, to extend the SDA by taking into account the
damping of the distribution function $\delta f$ , it is reasonable to
replace the $\delta $-functions in Eq. (\ref{dst}) by Lorentzians of the
form
\begin{equation}
\delta (\Delta \epsilon \pm \hbar \omega )\Longrightarrow \frac{1}{\pi }%
\frac{\hbar \omega _{I}}{(\Delta \epsilon \pm \hbar \omega _{R})^{2}+(\hbar
\omega _{I})^{2}}.  \label{delta1}
\end{equation}%
Using Eqs. (\ref{t1}), (\ref{dst}) and (\ref{delta1}), one then obtains, see
also Ref. \cite{kolu98},%
\begin{equation}
\tau =\frac{4\pi ^{2}\alpha \hbar }{R^{(+)}+R^{(-)}}\int_{V_{\mathrm{eq}%
}}^{\infty }d\epsilon (\epsilon -V_{\mathrm{eq}})^{3/2}(\partial f/\partial
\epsilon )_{\mathrm{eq}}  \label{ts2}
\end{equation}%
where
\begin{equation}
R^{(\pm )}=-\frac{3\hbar \omega _{I}}{2\pi }\int_{V_{\mathrm{eq}}}^{\infty }%
\frac{d\epsilon _{1}d\epsilon _{2}d\epsilon _{3}d\epsilon _{4}\ (\epsilon
_{1}-V_{\mathrm{eq}})^{3/2}}{(\Delta \epsilon \pm \hbar \omega
_{R})^{2}+(\hbar \omega _{I})^{2}}\left( \frac{5}{2}\frac{\delta Q}{\delta
f_{1}}\frac{\partial f_{1}}{\partial \epsilon _{1}}+\frac{1}{2}\frac{\delta Q%
}{\delta f_{2}}\frac{\partial f_{2}}{\partial \epsilon _{2}}+\frac{\delta Q}{%
\delta f_{3}}\frac{\partial f_{3}}{\partial \epsilon _{3}}\right) _{\mathrm{%
eq}}.  \label{r1}
\end{equation}%
Here and below we assume a constant spherical potential of radius $R_{0}$
for $V_{\mathrm{eq}}$\ and use the Fermi function for the equilibrium
distribution function $f_{\mathrm{eq}}$
\begin{equation}
f_{\mathrm{eq}}=\left[ 1+\exp \left( \frac{p^{2}/2m-\lambda }{T}\right) %
\right] ^{-1},  \label{feq1}
\end{equation}%
\begin{equation*}
\lambda \simeq \epsilon _{F}\left[ 1-\frac{\pi ^{2}}{12}\left( \frac{T}{%
\epsilon _{F}}\right) ^{2}\right] ,
\end{equation*}%
where $p^{2}/2m=\epsilon -V_{\mathrm{eq}},\ \epsilon _{F}=p_{F}^{2}/2m$ and $%
p_{F}$\ is the Fermi momentum

We will reduce the local equations (\ref{sys1})-(\ref{sys3}) to macroscopic
equations of motion for the nuclear shape variables assuming the following
separable form for the displacement field%
\begin{equation*}
\mathbf{\chi }(\mathbf{r},t)=\beta (t)\mathbf{v}(\mathbf{r}),
\end{equation*}%
where $\beta (t)=\beta _{\omega }\exp ({-i\omega t})$ is the collective
variable. Considering quadrupole vibrations, we will assume an irrotational
motion with \cite{risc80}
\begin{equation}
\mathbf{v}(\mathbf{r})=\mathbf{\nabla }(r^{2}Y_{20}(\hat{r}))/2.  \label{v1}
\end{equation}

Substituting Eq. (\ref{v1}) into Eq. (\ref{sys2}) we obtain the equation of
motion for the collective variable $\beta (t)$,
\begin{equation}
m\rho _{\mathrm{eq}}v_{\alpha }\ddot{\beta}+\left( \rho _{\mathrm{eq}}{\frac{%
\partial }{\partial r_{\alpha }}}\left( {\frac{\delta ^{2}\mathcal{E}}{%
\delta \rho ^{2}}}\right) _{\mathrm{eq}}\delta \rho _{\omega }+\frac{%
\partial \pi _{\nu \alpha ,\omega }}{\partial r_{\nu }}\right) \beta =0,
\label{eq.mov}
\end{equation}%
where%
\begin{equation}
\pi _{\nu \alpha ,\omega }=P_{\alpha \beta ,\omega }^{\prime }/\beta
_{\omega }.  \label{drhopi}
\end{equation}%
Multiplying Eq. (\ref{eq.mov}) by $v_{\alpha }$, summing over $\alpha $, and
integrating over $\mathbf{r}$-space, we obtain the dispersion equation for
the eigenfrequency $\omega $,
\begin{equation}
-B\omega ^{2}+C_{LD}+C^{\prime }(\omega )-i\omega A(\omega )=0.  \label{eq.w}
\end{equation}%
Here, $B$ is the mass coefficient with respect to the collective variable $%
\beta (t)$:
\begin{equation}
B=m\int d\mathbf{r\ }\rho _{\mathrm{eq}}(\mathbf{r})|\mathbf{v(r)}|^{2}.
\label{B}
\end{equation}%
The dissipative term $A(\omega )$ and the stiffness coefficient $C^{\prime
}(\omega )$ are caused by the dynamic distortion of the Fermi surface and
they are given by
\begin{equation}
A(\omega )=\int d\mathbf{r}\ \frac{1}{\omega _{R}}\ \mathrm{Im}\ \pi _{\nu
\alpha ,\omega }\ \nabla _{\nu }v_{\alpha },  \label{a1}
\end{equation}%
\begin{equation}
C^{\prime }(\omega )=\int d\mathbf{r}\left( \ \frac{\omega _{I}}{\omega _{R}}%
\ \mathrm{Im}\ \pi _{\nu \alpha ,\omega }-\mathrm{Re}\ \pi _{\nu \alpha
,\omega }\right) \nabla _{\nu }v_{\alpha }.  \label{c1}
\end{equation}

The stiffness coefficient $C_{LD}$ in Eq. (\ref{eq.w}) can be identified
with that of the traditional liquid drop model \cite{bomo2}
\begin{equation}
C_{LD}=\frac{1}{\pi }b_{S}A^{2/3}-\frac{1}{2\pi }b_{C}\frac{Z^{2}}{A^{1/3}},
\label{cldm}
\end{equation}%
where $b_{S}=17.2$\textrm{\ MeV} and $b_{C}=0.7$\textrm{\ MeV} are the
surface tension coefficient and the Coulomb energy coefficient,
respectively. To evaluate the kinetic coefficients $A(\omega ),$ $B$ and $%
C^{\prime }(\omega )$ we will assume a sharp surface of the nucleus in
equilibrium $\rho _{\mathrm{eq}}=\rho _{0}\theta (R_{0}-r)$, where $\rho _{0}
$\ is the bulk density and $R_{0}=r_{0}A^{1/3}$ \textrm{fm}. Taking into
account the expression (\ref{v1}) for the displacement field, we obtain from
Eq. (\ref{B}) the mass coefficient as
\begin{equation}
B=\frac{3}{8\pi }AmR_{0}^{2}.  \label{b2}
\end{equation}%
Using Eqs. (\ref{2.10}), (\ref{2.11}), (\ref{2.12}), (\ref{v1}) and (\ref%
{drhopi}), we obtain from Eqs. (\ref{a1}) and (\ref{c1}) the kinetic
coefficients $A(\omega )$ and $C^{\prime }(\omega )$ in the following form%
\begin{equation}
A(\omega )=\frac{\tau }{(1-\omega _{I}\tau )^{2}+(\omega _{R}\tau )^{2}}\
\mathcal{J},  \label{a2}
\end{equation}%
\begin{equation}
C^{\prime }(\omega )=\frac{(\omega _{R}\tau )^{2}+(\omega _{I}\tau )^{2}}{%
(1-\omega _{I}\tau )^{2}+(\omega _{R}\tau )^{2}}\ \mathcal{J}.  \label{c2}
\end{equation}%
Here,
\begin{equation}
\mathcal{J}=\ \int d\mathbf{r}P_{\mathrm{eq}}\ \Lambda _{\alpha \beta }\
\nabla _{\beta }v_{\alpha }=5R_{0}^{3}P_{0}  \label{j1}
\end{equation}%
and
\begin{equation*}
P_{0}={\frac{1}{5}}{\frac{\hbar ^{2}}{m}}\left( {\frac{6\pi ^{2}}{g}}\right)
^{2/3}\rho _{0}^{5/3}.
\end{equation*}

\section{Numerical calculations}

For the numerical calculations in this work we adopt the value of $%
r_{0}=1.12\,\mathrm{fm.}$ In \textrm{Fig. 1}, we have plotted the results of
the calculations of the excitation energy $E=\hbar \omega _{R}$ and the
collisional width $\Gamma =2\hbar \omega _{I}$ obtained from Eq. (\ref{eq.w}%
) for the nucleus with $A=224$ and $Z=A[1-2A/5(A+200)]/2=88$, which
corresponds to the valley of beta-stability \cite{risc80}. We have here
adopted the value of $\alpha =9.2$ \textrm{MeV} from \cite{kopl95}, which
corresponds to $\sigma _{\mathrm{in}}\approx \sigma _{\mathrm{free}}/2,$
where $\sigma _{\mathrm{free}}\approx 40$ \textrm{mb} is the cross section
for the nucleon-nucleon scattering in free space. The eigenenergy $E$\ in a
cold nucleus is strongly shifted up with respect to the one in the nuclear
liquid drop model because of the contribution from the term $C^{\prime
}(\omega )$ to the nuclear stiffness coefficient $C=C_{LD}+C^{\prime
}(\omega )$\ in Eq. (\ref{eq.w}). The stiffness coefficient $C^{\prime
}(\omega )$ is caused by the dynamic Fermi surface distortion effect and it
decreases strongly with increasing temperature $T$. Due to this fact, the
stiffness coefficient $C$ and the eigenenergy $E$ decrease monotonously with
temperature and approach the liquid drop model limit for large temperatures,
see also Ref. \cite{kiko96}. As seen from \textrm{Fig. 1}, the SDA leads to
a significantly faster decrease of $E$ with increasing temperature $T$,
i.e., the SDA overestimates the decrease rate of the Fermi surface
distortion effect with temperature. The damping of $\delta f$ in the
collision integral does not play an essential role in the derivation of the
collisional width $\Gamma $ of the GQR. We can see from \textrm{Fig. 1} that
the damping of $\delta f$ increases the width $\Gamma $ with respect to the
SDA, but the additional contribution to the width does not exceed $20\%$.

The difference between the SDA and its extension of Eq. (\ref{ts2}) becomes
stronger for smaller values of the parameter $\alpha $. Let us introduce the
ratio%
\begin{equation}
q=\tau _{\mathrm{SDA}}/\tau ,  \label{q}
\end{equation}%
which allows one to estimate the value of the correction for the two-body
relaxation time due to the damping of the distribution function $\delta f$
in the collision integral. The result of the calculations of $q$ as a
function of the temperature $T$,\ with $\omega _{R}$ and $\omega _{I}$ taken
from \textrm{Fig.~1}, is shown in \textrm{Fig.~2} for two values of
parameter $\alpha =9.2$ \textrm{MeV} and $\alpha =4.6$ \textrm{MeV}. (Note
that the relaxation time $\tau $\ is a complicated function of $\alpha $\
because the quantities $\omega _{R}$\ and $\omega _{I}$\ in Eq. (\ref{r1})
depend on $\alpha $ due to Eqs. (\ref{eq.w}), (\ref{a2}) and (\ref{c2}).) In
\textrm{Fig.~3 }we show the temperature dependence of the ratio
\begin{equation*}
\Delta \Gamma /\Gamma =(\Gamma -\Gamma _{\mathrm{SDA}})/\Gamma ,
\end{equation*}%
for two values of the parameter $\alpha $. Here $\Gamma =\hbar \omega _{I}$
is the exact collisional width obtained using the relaxation time $\tau $\
of Eq. (\ref{ts2}) and $\Gamma _{\mathrm{SDA}}$ is the one obtained within
the SDA using $\tau _{\mathrm{SDA}}$\ from Eq. (\ref{t2}).\ The
non-monotonic behavior of the ratio $\Delta \Gamma /\Gamma $\ in the case of
$\alpha =4.6$ \textrm{MeV} (dashed line) is because the width $\Gamma _{%
\mathrm{SDA}}$ has a bell-shaped form as a function of $T$, for moderate
temperatures \cite{kopl95}.\ \

\section{Conclusions}

In this paper we have studied the validity of the small damping
approximation for the quasi-classical description of the averaged properties
of nuclei at high temperatures in the framework of the Fermi liquid theory.
The basic equation of the collisional kinetic theory has been reduced to
macroscopic equations of motion for the shape variable of the nucleus. We
have extended the small damping approximation\ taking into account the
damping of the distribution function $\delta f$ in the collision integral $%
\delta \mathrm{St}[\delta f]$ and replacing the energy conservation $\delta $%
-functions in $\delta \mathrm{St}[\delta f]$ by Lorentzians. Such kind of an
extension of the SDA leads to an increase in of the relaxation rate. The
corresponding decrease of the collisional relaxation time $\tau $\ does not
exceed $20\%$ of the value $\tau _{\mathrm{SDA}}$ obtained within the SDA
and it decreases with increasing temperature $T$.

We have shown that the extension of the SDA due to the damping of $\delta f$
in the collision integral reduces the rate of variation with temperature of
the Fermi surface distortion effects. The smoothing of the $\delta $-function
in the collision integral of Eq. (\ref{dst}) caused by the damping of the
distribution function $\delta f$ provides much more weaker dependence of
eigenenergy $E$ on the temperature than the one in the SDA.

The difference between the SDA and its extension of Eq. (\ref{ts2}) becomes
stronger for smaller values of the relaxation parameter $\alpha $. The
dashed line in \textrm{Fig. 3} shows that the correction to the SDA due to
the damping of the $\delta f$ in the collision integral can increase the
collisional width $\Gamma $ of the giant quadrupole resonance for a small
enough value of the parameter $\alpha $.

\bigskip

\section{Acknowledgments}

This work was supported in part by the US Department of Energy under grant
\# DOE-FG03-93ER40773 and the US National Science Foundation under grant \#
0355200. One of us (V.M.K.) acknowledges the nice hospitality of the
Cyclotron Institute at Texas A\&M University.

\medskip \newpage

\bigskip

\bigskip

\begin{center}
\textbf{Figure captions}
\end{center}

\medskip

Fig. 1. Temperature dependence of the collisional width $\Gamma =2\hbar
\omega _{I}$ and excitation energy $E=\hbar \omega _{R}$ of the GQR for the
nucleus with mass number $A=224$ obtained from Eq. (\ref{eq.w}) with $\alpha
=9.2$ \textrm{MeV}. The solid lines are obtained using $\tau $\ from Eq. (%
\ref{ts2}) and the dashed lines are for the SDA with Eq. (\ref{t2}).

Fig. 2. The dependence of the ratio $q$ for the GQR for the nucleus with
mass number $A=224$ on the temperature $T$ obtained from the solution of Eq.
(\ref{eq.w}). The solid and dashed lines correspond to the calculation with
the parameter $\alpha =9.2$ \textrm{MeV} and $\alpha =4.6$\textrm{\ MeV},
respectively.

Fig. 3. Temperature dependence of the relative correction $\Delta \Gamma
/\Gamma $ to the width of the GQR due to the damping effect in the collision
integral for the nucleus with mass number $A=224$. The solid and dashed
lines correspond to the calculation with the parameter $\alpha =9.2$\textrm{%
\ MeV} and $\alpha =4.6$ \textrm{MeV}, respectively.

\bigskip

\bigskip

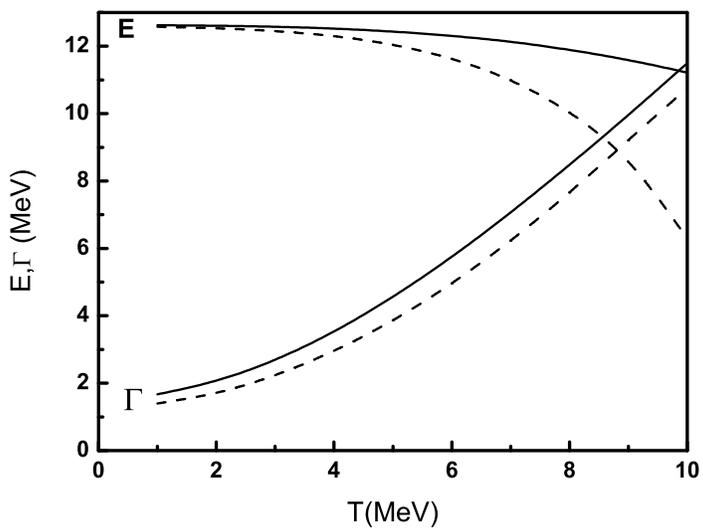
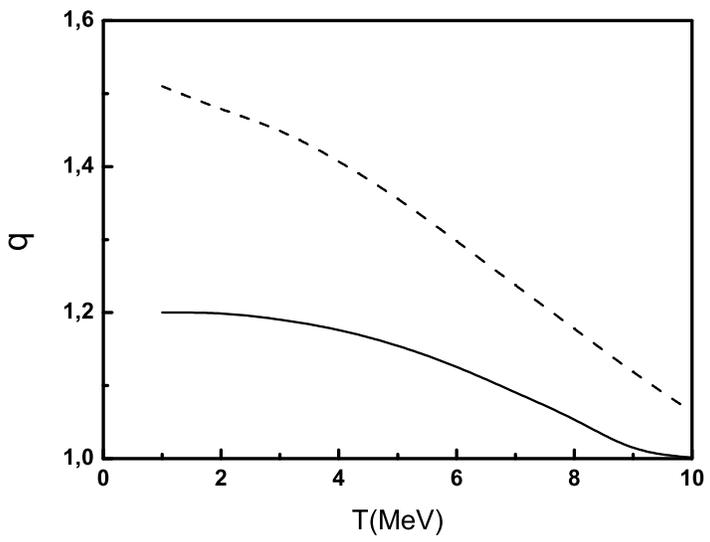
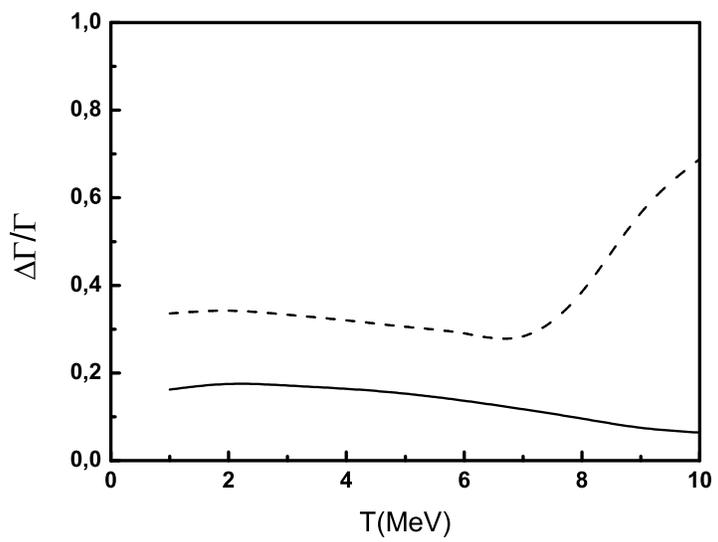
\end{document}